\documentclass[reprint,aps,prl,amsmath,amssymb,showpacs,superscriptaddress,]{revtex4-1}
\usepackage{graphicx}
\usepackage{dcolumn}
\usepackage{bm}
\usepackage{color}
\usepackage{upgreek}

\begin{document}

\title{Enhanced high-order harmonic generation from periodic potentials in inhomogeneous laser fields}

\author{Tao-Yuan Du }\affiliation{State Key Laboratory of Magnetic Resonance and Atomic and Molecular Physics, Wuhan Institute of Physics and Mathematics, Chinese Academy of Sciences, Wuhan 430071, China}\affiliation{University of Chinese Academy of Sciences, Beijing 100049, China}

\author{Zhong Guan}\affiliation{State Key Laboratory of Magnetic Resonance and Atomic and Molecular Physics, Wuhan Institute of Physics and Mathematics, Chinese Academy of Sciences, Wuhan 430071, China}\affiliation{College of Physics and Electronic engineering, Northwest Normal University, Lanzhou 730070, China}

\author{Xiao-Xin Zhou}\affiliation{College of Physics and Electronic engineering, Northwest Normal University, Lanzhou 730070, China}

\author{Xue-Bin Bian}\email{xuebin.bian@wipm.ac.cn}\affiliation{State Key Laboratory of Magnetic Resonance and Atomic and Molecular Physics, Wuhan Institute of Physics and Mathematics, Chinese Academy of Sciences, Wuhan 430071, China}

\begin{abstract}

We theoretically study high-order harmonic generation (HHG) from solid-phase systems in spatially inhomogeneous strong laser fields originated by resonant plasmons within a metallic nanostructure. The intensity of the second plateau in HHG may be enhanced by two-three orders and be comparable with the intensity of the first plateau. This is due to bigger transition probabilities to higher conduction bands. It provides us a practical way to increase the yields of HHG with laser intensity below the damage threshold. It presents a promising way to triple the range of HHG spectra in experimental measurements. It also allows to generate intense isolated attosecond pulse from solids driven by few-cycle laser fields.


\end{abstract}

\maketitle

The study of light-matter interaction with intense laser fields is a rapidly growing research area, which exhibits many novel phenomena \cite{Corkum, Pronin1, Pronin2}. The high-order harmonic generation (HHG) has attracted a lot of attention since it provides a tabletop coherent x-ray source. It has been used for dynamic imaging of molecular structures \cite{Itatani, Bian, Smirnova}. However, the low conversion efficiency of HHG from gas-phase systems restricts its applications.
Recently, HHG has been experimentally generated from bulk crystals \cite{Ghimire, Schubert, Luu}. Due to high density of solid-state materials, it is possible to generate HHG with higher conversion efficiency and to probe the structure of solids. Vampal \textit{et al}. realized all-optical reconstruction of crystal band structure \cite{Vampa1}. This method extends measurement schemes of solid-state material band structures.
Theoretically, HHG in solid-state materials involves two contributions (inter-band and  intra-band currents). The experimental measurement \cite{Ghimire} shows one main plateau in HHG. However, the theoretical simulations \cite{Wu, Guan} reveal double-plateau structures. The primary plateau is a result of the resonant transition between the valence band and the first conduction band, while a weaker second plateau is due to transitions from high-lying conduction bands. The intensity of the second plateau is a few orders lower than that of the first one. This may be the reason why the second plateau has not been well resolved experimentally.
 Recently, an alternative technique, named the spatially inhomogeneous field (or source of plasmonic enhancement), assisted generation of ultrashort attosecond (1 as = 10$^{-18}$ s) sources in atoms and molecules, which has attracted much attention both in experiments and theories \cite{Kim1, Sivis, Shaaran1, Shaaran2, Shaaran3, Ciappina1, Ciappina2, Yavuz, He, Cao, Park, Husakou, Perez-Hernandez, Wang}. In comparison with the homogeneous field, the electrons obtain extra energies by further acceleration in the inhomogeneous fields. This alternative technique provides a way to increase the effective intensity of the input laser fields by almost three orders. As we know, the HHG from solids suffers from low damage-threshold intensity. This plasmon-enhanced scheme may overcome this shortage. In this work, we will theoretically study the HHG from solids in inhomogeneous laser fields. To our knowledge, this has not been investigated.

Based on the single-electron approach, we describe the light-solid interaction in one dimension, along the polarization direction of the laser fields. In the length-gauge treatment, the time-dependent Hamiltonian is written as
\begin{equation}\label{E1}
\hat{H}(t)=\hat{H}_0 + exE(x,t),
\end{equation}
where $\hat{H}_0=\frac{{\hat{p}^2}}{2m}+V(x)$, and $V(x)$ is a periodic lattice potential. In our calculations, we choose the Mathieu-type potential \cite{Slater}. The specific form is $V(x)=-V_0[1+\cos(2\pi x/a_0)]$, with $V_0=0.37$ a.u. and lattice constant $a_0=8$ a.u. The spatial dependence of the enhanced laser electric field is perturbative and linear with respect to the position, and the laser field can be approximated as
\begin{equation}\label{E2}
  E(x,t) \simeq E(t)(1 + \epsilon x ),
\end{equation}
where $\epsilon  \ll 1$ is a parameter characterizing the strength of inhomogeneity. The time-dependent Hamiltonian can be rewritten as
 \begin{equation}\label{E3}
\hat{H}(t)=\hat{H}_0 + exE(t) + e(\epsilon x^2)E(t),
\end{equation}
where the first two terms account for the Hamiltonian in the case of homogeneous fields and the last term is an additional term induced by the inhomogeneous field.
In Fig. 1, we show a schematic illustration of the laser field enhancement using a nanostructure of bow-tie elements. The parameter \emph{d} and the shape of nanostructure adjust the field inhomogeneity parameter $\epsilon$.
In absence of the laser fields, the time-independent Schr\"odinger equation (TDSE) can be written as

\begin{equation}\label{E4}
\hat{H_0}\phi_{n}(x)=E_{n}\phi_{n}(x).
\end{equation}

We use \emph{B}-spline functions \cite{Bachau} to expand the time-independent wave function,

\begin{equation}\label{E5}
\phi_{n}(x)=\sum_{i=1}^{N_{max}}c_{i}B_{i}(x).
\end{equation}

Substituting Eq. (\ref{E5}) into Eq. (\ref{E4}), we obtain matrix equation

\begin{equation}
HC=ESC,
\end{equation}
where $C$ is the column matrix, $H$ and $S$ are $N\times N$ square matrices, respectively.

We use 2400 \emph{B}-splines to calculate its eigenvalues in the space region [-240, 240] a.u. and obtain the energy bands structure which agrees well with the calculations by Bloch-state expansion. By using \emph{B}-spline basis, we perform all the calculations in the coordinate space. For the details, we refer readers to Ref. \cite{Guan}. Due to the drive of laser fields, electrons in the valence band have probabilities to tunnel to conduction bands, i.e. Zener tunneling. But the tunneling probabilities exponentially decay with the increase of energy gap. Only a small portions of populated electrons near the wave-vector $k=0$ on top of the valence band can tunnel to conduction bands with the laser parameters used in the current work. So we choose an initial state calculated by our \emph{B}-spline method with $k=0$ on top of the valence band with the minimum band gap.
We can effectively solve the time-dependent $|\psi{}(t)\rangle$ by using symmetrically split-operator algorithm \cite{Feit}

\begin{equation}\begin{split}
|\psi{}(x,t_0+\Delta t)\rangle &=e^{-i\hat{T}\frac{\Delta t}{2}} e^{-i\hat{V}_{eff} \Delta t}e^{-i\hat{T}\frac{\Delta t}{2}}|\psi{}(x,t_0)\rangle
\\& + O(\Delta t^{3}),\end{split}\end{equation}
where $\hat{V}_{eff}=V(x)+exE(x,t)$ and $\hat{T}$ is the kinetic operator.

In our calculations, we use the laser pulses with a $\cos^2$ envelope and the grid spacing is 0.03 a.u., which is sufficient to obtain converged results. The time step is 1/4096 of an optical cycle. In order to prevent spurious reflections from the boundary, the total wave function is multiplied by a mask function of the form cos$^{1/8}$ with $|x|>216$ a.u. at each time step.
After obtaining the time-dependent $|\psi{}(t)\rangle$ at an arbitrary time, we can calculate the time-dependent laser-induced currents by

\begin{equation}
j(t)=-\frac{e}{m}\left[\langle \psi{}(t)|\hat{p}|\psi{}(t)\rangle\right].
\end{equation}

The HHG power spectrum is proportional to the modulus square $|\emph{j}(\omega)|^{2}$ of Fourier transform of the time-dependent current in Eq. (8). Before the Fourier transform, we multiply \emph{j}(t) by a time-dependent Hanning window function in order to suppress the dipole moment due to the population remaining in the conduction bands at the end of the pulses.
Based on solid-physics theory, the energy eigenvalues of solid-state materials show a multiple-band structure. Each band group can be distinguished easily as illustrated in Ref. \cite{Guan}. The intra-band contribution to the current is induced by transitions within the same band, while the inter-band contribution mainly involves transitions between the conduction bands and valence band. By distinguishing the contributions from intra- or inter-band, we elucidate the mechanism of intensity enhancement in the second HHG plateau.

\begin{figure}
\centering\includegraphics[width=7.5 cm,height=3.5 cm]{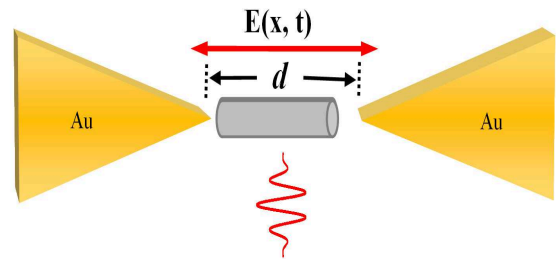}
\caption{(Color online) Schematic illustration of laser field enhancement using a nanostructure of bow-tie elements. The cylinder represents solid-state material samples. }\label{Fig1}
\end{figure}

First, we study the harmonic spectra of the periodic systems under a mid-infrared laser pulses. The peak intensity and the wavelength of the driving laser pulses are $I$ = 8.774$\times 10^{11}$ W/cm$^2$ and $\lambda$ = 3200 nm, respectively. We show the harmonic spectra in the case of the homogeneous and the inhomogeneous fields with different inhomogeneity parameters $\epsilon$ in Fig. 2(a). The harmonic spectra show a two-plateau structure in both the homogeneous field and the inhomogeneous field. However, the second HHG plateau exhibits two-three orders intensity enhancement in the inhomogeneous fields with $\epsilon = 0.0005$ compared with the case in the homogeneous field. In order to distinguish the physics behind the harmonic spectra, we show the first HHG plateau in Fig. 2(b), and the second HHG plateau in Fig. 2(c). One can clearly observe both the odd and even order harmonics from the HHG spectra in the case of inhomogeneous field. The generation of the even harmonics is due to the fact that the symmetry of the system is broken. The energy of the first few fast decaying HHG is lower than the band gap. They show a perturbative character which is similar to the below-threshold HHG in the gas phase.

\begin{figure}
\centering\includegraphics[width=8 cm,height=6.5 cm]{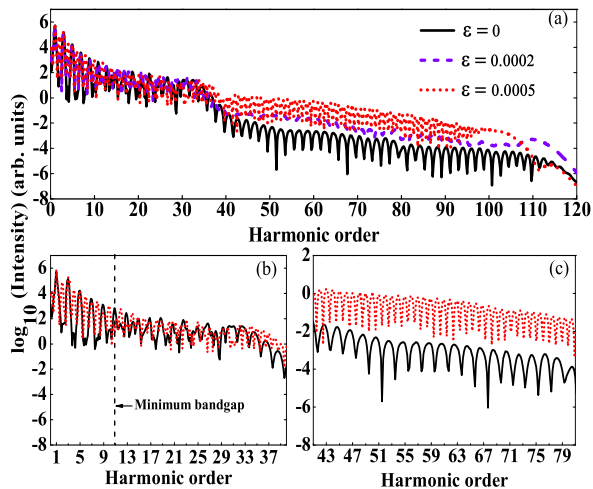}
\caption{(Color online) HHG spectra in the case of the homogeneous field (black solid curve), the inhomogeneous fields with $\epsilon$ = 0.0002 (dash violet curve) and $\epsilon$ = 0.0005 (red dot curve). (a) HHG with full range. (b) The first HHG plateau with $\epsilon$ = 0.0005. (c) The second HHG plateau with $\epsilon$ = 0.0005. The intensity, wavelength and duration of the driving laser pulses are $8.77\times 10^{11}$ W/cm$^{2}$, 3200 nm and six-cycle, respectively. }\label{Fig2}
\end{figure}

  In order to obtain further insights into the effect of spatial inhomogeneity on the HHG process in Fig. 2, we make a distinction about HHG induced by conduction band C1 and C2 plus C3 contributions in Fig. 3, respectively. The mechanism of the first HHG plateau is mainly determined by inter-band transitions between the conduction band C1 and the valence band, which have been pointed out by previous researches in homogeneous field \cite{Vampa1,Wu,Guan}. One can find that the second HHG plateau is the results of inter-band transitions between conduction bands C2 plus C3 and the valence band.

\begin{figure}
\centering\includegraphics[width=8 cm,height=4.5 cm]{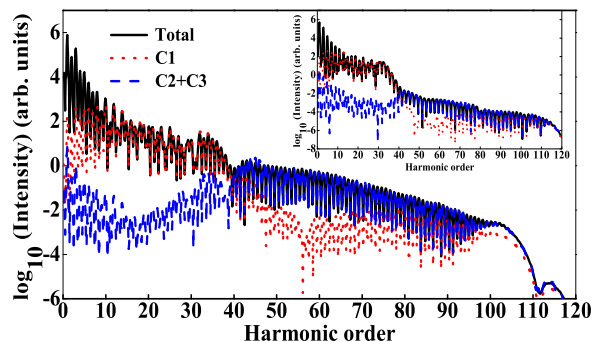}
\caption{(Color online) HHG induced by full (black solid curve), conduction band C1(red dot curve) and C2 plus C3(blue dash curve) contributions in an inhomogeneous field with $\epsilon$ = 0.0005. The inset shows the HHG induced by full, conduction band C1 and C2 plus C3 contributions in the homogeneous field. The laser pulse parameters are the same as those in Fig. 2.}\label{Fig3}
\end{figure}

   We then perform a time-frequency analysis \cite{Chandre} of the harmonics obtained in Fig. 2, as shown in Fig. 4(a) and (b), respectively. However, the harmonic emission behavior changes dramatically in the case of inhomogeneous field in Fig. 4(b), which disturbs the HHG trajectories and generates both odd and even-order harmonic spectra. For the second HHG plateau, the spatial inhomogeneity dramatically affects the harmonic emission process, which presents enhancement processes around the -0.25 o.c. and 0.75 o.c. and shows a suppressed process at 0.25 o.c. In order to further explain the phenomenon that only the second HHG plateau has a two-three orders magnitude enhancement in the case of inhomogeneous field, we present the time-frequency analysis of the inter-band contributions in Fig. 4(c) and (d). It clearly shows a two-three orders magnitude enhancement of the second HHG plateau, while the magnitude of the first HHG plateau is still comparable. This agrees with our analysis \cite{Guan} that interband transitions play key roles in HHG and thus may explain why only the second plateau is dramatically enhanced.

\begin{figure}
\centering\includegraphics[width=8 cm,height=6.5 cm]{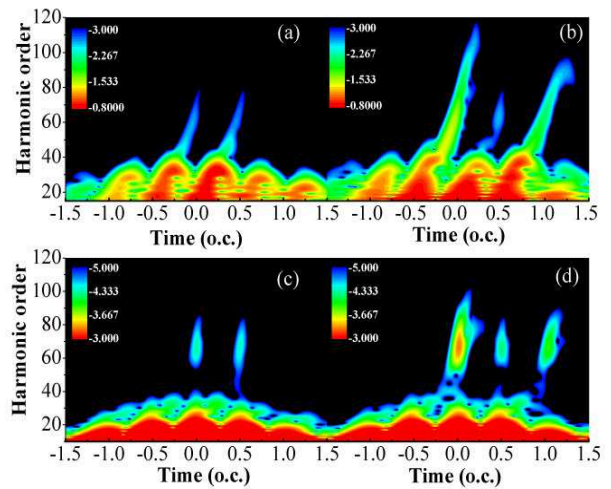}
\caption{(Color online) Time-Frequency analysis of HHG in a homogeneous field in (a), (c) and an inhomogeneous field  in (b), (d). (a)-(b) are time-frequency analysis with full contribution. Due to the fact that the two HHG plateaus are mainly determined by inter-band contribution, we show time-frequency analysis of inter-band contribution in homogeneous field in (c) and inhomogeneous field in (d).}\label{Fig4}
\end{figure}

  We finally show energy band structure and time-dependent electron population of the conduction bands in Fig. 5. In Fig. 5(a) and (b), one can find that the energy bands calculated by Bloch states agree well with those by \emph{B}-spline basis. The minimum gap between the valence band and the first conduction band is about 4.2 eV, which corresponds to harmonic order 11 in Fig. 2(b). According to the energy band structure and the population of the conduction bands, one can further get insight into the HHG enhancement. As previously mentioned, the harmonic plateaus are determined by the transition process between the conduction bands and valence band, i.e. inter-band contribution. Transition between the first conduction band and valence band contributes to the first plateau, while the second plateau is attributed to transitions between high-lying conduction bands and the valence band. In Fig. 5(c), it illustrates that the first conduction band electron population has a small increment in the case of inhomogeneous field compared with that in the case of homogeneous field, which explains why the change of the intensity of the first plateau is not obvious. However, in Fig. 5(d), One can clearly see that electron population of the second conduction band has a dramatic increment at the center of the laser pulse, which could lead to more electrons coupling to valence band and give rise to  two-three orders enhancement of the second HHG plateau.

\begin{figure}
\centering\includegraphics[width=8.5 cm,height=7 cm]{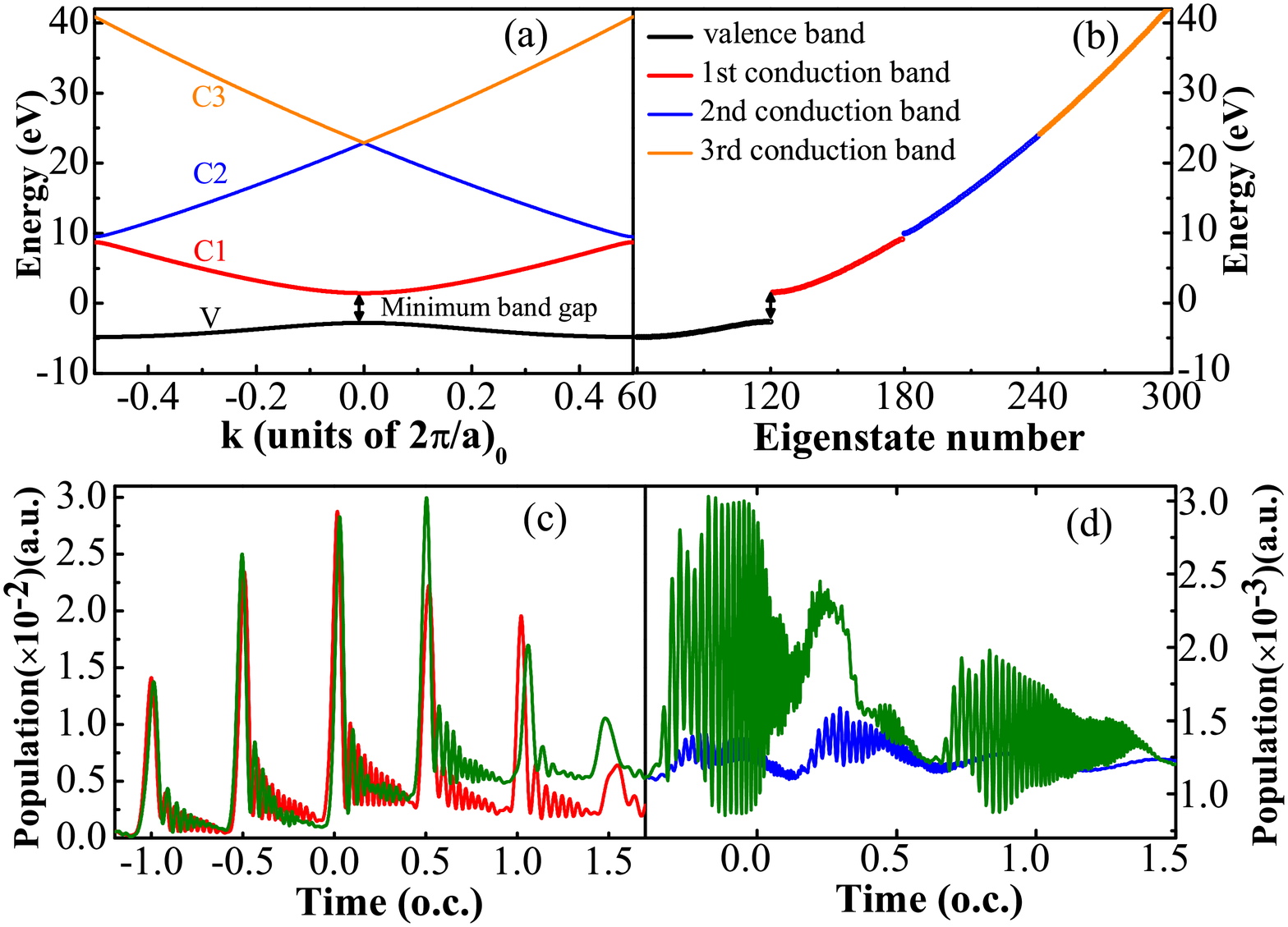}
\caption{(Color online) Band structures calculated by Bloch states expansion compared with the result obtained by \emph{B}-spline basis in (a) and (b), respectively. The electron population as a function of time for the 1st conduction band (red curve) and the 2rd conduction band (blue curve) in homogeneous field are illustrated in (c) and (d), respectively. The olive curves present results in the inhomogeneous field. }\label{Fig5}
\end{figure}

  In addition, we calculate the transition probabilities between the conduction and the valence bands in Fig. 6, to reveal the population enhancement in the case of inhomogeneous fields. The effect of spatial inhomogeneity on the transition probabilities can be described by the additional term $|\langle\phi_{n_0}|x^{2}|\phi_{n}\rangle|^2$, in which $n_0$ represents the initial valence band state and $n$ represents the conduction band states. One can clearly see two minimum values of transition probabilities in the case of homogeneous field (black solid curve) at the energy about 10 eV and 24 eV, which correspond to the two energy gap positions within the conduction bands in Fig. 5(a) and (b). The enhancement of transition probability is negligible in the first conduction band (corresponds to the energy range below 10 eV), while a one-two orders enhancement of transition probability has been obtained in the second conduction band (corresponds to the energy range from 10 eV to 24 eV) caused by the effect of spatial inhomoneity. This provides a deeper insight to the fact that one can just observe obvious enhancement of intensity in the second HHG plateau.

\begin{figure}
\centering\includegraphics[width=7.5 cm,height=6 cm]{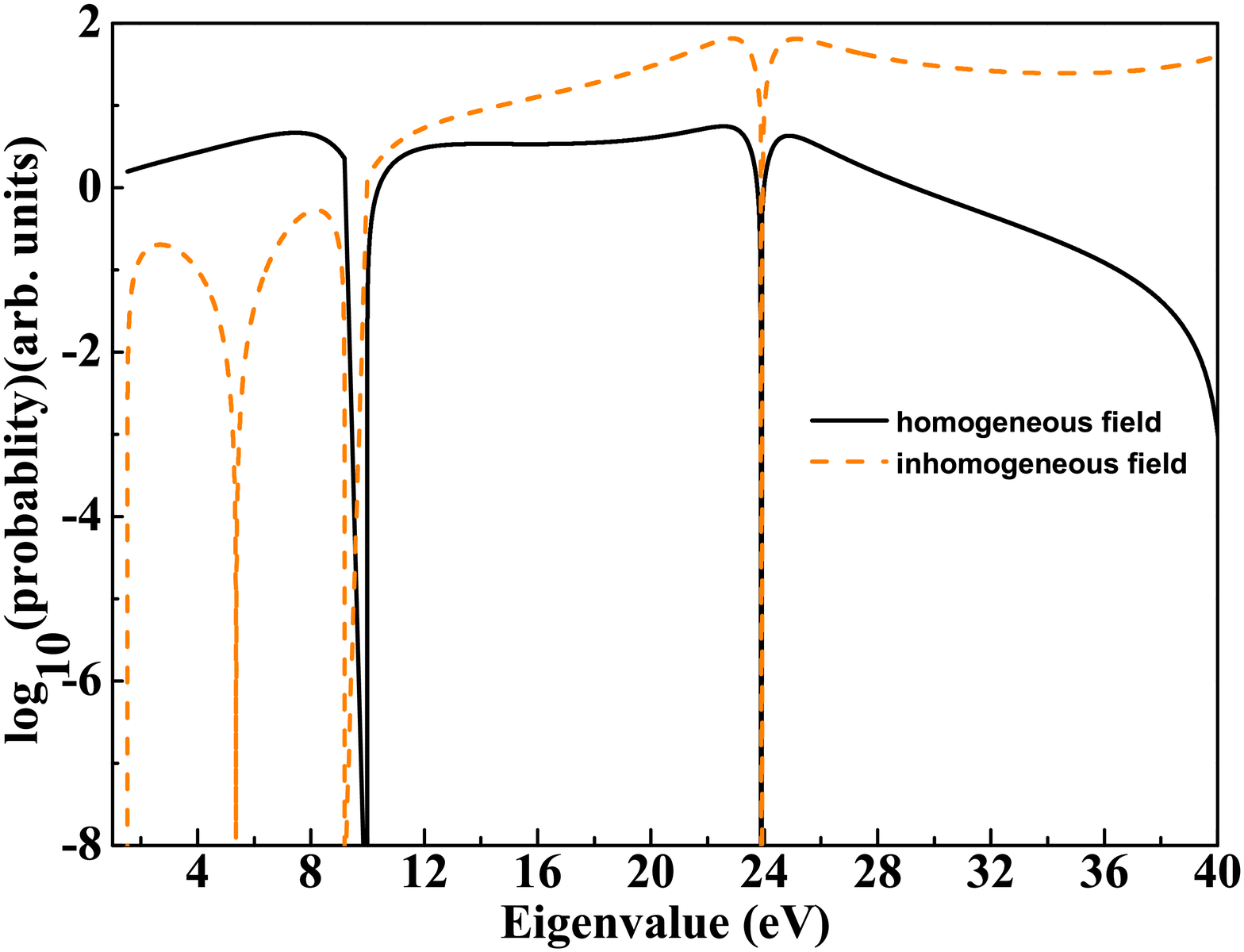}
\caption{(Color online) Transitional probabilities calculated by $|\langle\phi_{n_0}|\hat{x}|\phi_{n}\rangle|^2$ (black solid curve) for homogeneous field and an additional term $|\langle\phi_{n_0}|x^{2}|\phi_{n}\rangle|^2$ (orange dash curve) for spatial inhomogeneous field. }\label{Fig6}
\end{figure}

\begin{figure}
\centering\includegraphics[width=7.5 cm,height=7.5 cm]{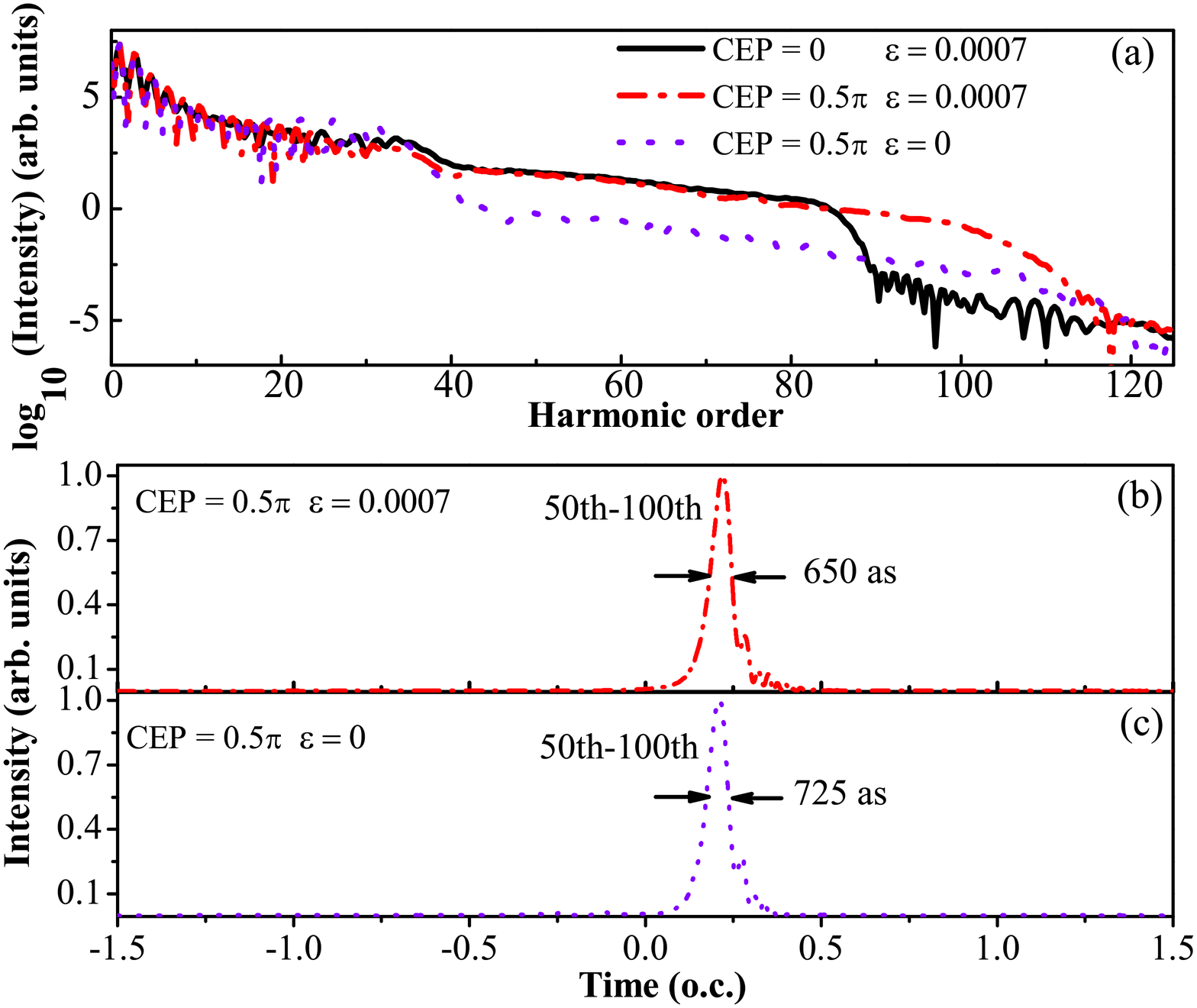}
\caption{(Color online) (a) HHG in few-cycle laser fields. For the inhomogeneous field, $\epsilon$ = 0.0007. We choose two different CEP values, 0 and 0.5$\pi$. The violet dot curve in (a) shows the HHG in the case of homogeneous field with CEP = 0.5$\pi$.  The field parameters are the same as those in Fig. 2 except for the pulse duration. Attosecond pulses generated by synthesis of harmonics with order 50-100 in the case of inhomogeneous field with $\epsilon$ = 0.0007, CEP = 0.5$\pi$ and in the case of homogeneous field with CEP = 0.5$\pi$ are presented in (b) and (c), respectively. The intensity has been normalized. }\label{Fig7}
\end{figure}

  Due to the intensity enhancement of the second HHG plateau in the case of inhomogeneous field, it can be used for intense isolated attosecond pulse generation. In order to obtain a super-continuous HHG spectra, we adopt three-cycle-duration pulses with inhomogeneity parameter $\epsilon$ = 0.0007. In Fig. 7(a), one can also observe two-three orders enhancement of intensity at the second HHG plateau in few-cycle pulses. The carrier-envelope phase (CEP) can be used to control the cutoff energy of HHG under the inhomogeneous fields. In Fig. 7 (b) and (c), the coherent superposition of harmonics has been adopted in the case of inhomogeneous field with $\epsilon$ = 0.0007 and homogeneous field with CEP = 0.5$\pi$. Isolated attosecond pulses with durations of 650 as and 725 as are produced by superposing the harmonics with order 50-100. One can draw a conclusion that the spatially inhomogeneous fields can not only realize two-three orders intensity enhancement of the second plateau but also control the HHG trajectories and narrow the width of the attosecond pulses.

  In summary, by numerically solving the TDSE, the HHG process in the solid-state materials under the action of a spatially inhomogeneous laser field is simulated in the coordinate space. The spatially inhomogeneous field has been demonstrated to be capable of realizing the dynamic control of quantum pathes. Two-band model is inadequate in this case. Multiband transitions and nonadiabatic effects have to be considered. As a result, a two-three orders intensity enhancement of the second plateau can be obtained in the case of inhomogeneous field, which is comparable with the intensity of the first plateau. This attributes to bigger transition probabilities to the high-lying conduction bands as a result of the enhancement of the transition matrix elements. An efficient scheme to enhance HHG yields below the damage threshold has been proposed. In the homogeneous laser field, the intensity of the second plateau is quite low. Only the intense first plateau in HHG is experimentally observed. This work sheds light upon enhancement of the second HHG plateau and a promising way to triple the range of HHG spectra in experimental measurements. This can also be used to generate intense narrow isolated attosecond pulses from solids.

The authors thank Mu-Zi Li and Xin-Qiang Wang very much for helpful discussions. This work is supported by the National Natural Science Foundation of China (Grants No. 11404376, No. 11561121002 and No. 11465016).

\end{document}